\documentclass[referee]{aa}  

\usepackage{graphicx}
\usepackage{txfonts}
\usepackage{graphics,graphicx,rotating,amsmath,color}

\newcommand{\bbe}{{\boldsymbol \beta}}
\newcommand{\bth}{{\boldsymbol \theta}}
\newcommand{\bde}{{\boldsymbol \delta}}
\newcommand{\bep}{{\boldsymbol \epsilon}}
\newcommand{\bepZ}{{\bep_{0th}}}

\newcommand{\mbe}{{\mbox{\boldmath$e$}}}
\newcommand{\mbeZ}{{\mbe_{0th}}}
\newcommand{\mbg}{{\mbox{\boldmath$g$}}}
\newcommand{\mbk}{{\mbox{\boldmath$k$}}}
\newcommand{\Real}[1]{{\rm Re}\left[ #1 \right]}
\newcommand{\rSN}{{\rm SN}}

\newcommand{\cZ}{{\cal Z}}

\newcommand{\lr}[1]{\left( #1 \right)}
\newcommand{\lrs}[1]{\left[ #1 \right]}

\newcommand{\IZt}{{I^{0}(\bbe)}}
\newcommand{\ILnsd}{{I^{Lnsd}}}
\newcommand{\ILnsdt}{{I^{Lnsd}(\bth)}}
\newcommand{\ISmd}{{I^{Smd}}}
\newcommand{\ISmdt}{{I^{Smd}(\bth)}}

\newcommand{\IESmd}{{I^{ESmd}}}

\newcommand{\hILnsdk}{{\hat I^{Lnsd}(\mbk)}}

\newcommand{\hISmdk}{{\hat I^{Smd}(\mbk)}}

\newcommand{\hIESmdk}{{\hat I^{ESmd}(\mbk)}}

\begin{document}

\title{A new weak lensing shear analysis method using ellipticity defined by 0th order moments}
\author{Yuki Okura\inst{1},Toshifumi Futamase\inst{2} }
\institute{
National Astronomical Observatory of Japan, Tokyo 181-8588, Japan\\
\email{yuki.okura@nao.ac.jp}\and
Astronomical Institute, Tohoku University, Sendai 980-8578, Japan\\
\email{tof@astr.tohoku.ac.jp}
}

%----------------------------------------------------------------------------------------------------
%----------------------------------------------------------------------------------------------------
%----------------------------------------------------------------------------------------------------
\abstract{
We developed a new method that uses ellipticity defined by 0th order moments (0th-ellipticity) for weak gravitational lensing shear analysis.
Although there is a strong correlation between the ellipticity calculated using this approach and the usual ellipticity defined by the 2nd order moment, 
the ellipticity calculated here has a higher signal-to-noise ratio because it is weighted to the central region of the image. These results were confirmed using data for Abell 1689 from the Subaru telescope.
For shear analysis, we adopted the ellipticity of re-smeared artificial image (ERA) method for point spread function (PSF) correction, and we tested the precision of this 0th-ellipticity with simple simulation, then we obtained the same level of precision with the results of ellipticity defined by quadrupole moments.
Thus, we can expect that weak lensing analysis using  0 shear will be improved in proportion to the statistical error.
}

\keywords{weak lensing}

\maketitle

%----------------------------------------------------------------------------------------------------
%----------------------------------------------------------------------------------------------------
%----------------------------------------------------------------------------------------------------
\section{Introduction}
Weak gravitational lensing has been widely recognized as a unique and very powerful method for studying not
only the mass distribution of the universe, but also  cosmological parameters (Mellier 1999, Schneider 2006, Munshi et al. 2008). 
In particular,  weak lensing studies have successfully revealed the mass distribution of the main halo of clusters of galaxies as well as 
sub-halos inside the clusters, and provide direct tests for the CDM structure formation scenario. 
Cosmic shear, namely weak lensing by large scale structure, has also attracted much attention recently because it may 
provide a way to measure dark energy, which may be responsible for the accelerated expansion of the universe.  
Although the signals due to cosmic shear have been measured by several independent groups
(Bacon et al 2000; Maoli et al 2001; Refregier et al 2002; Bacon et al 2003; Hamana et al 2003; Casertano et al 2003; van Waerbeke et al 2005; Massey et al 2005; Hoekstra et al 2006), much more accurate measurements are necessary 
to provide meaningful information on the nature of dark energy. 
This is extremely difficult because of the weakness of the signal, as well as  noise and uncontrollable systematic bias.  

In spite of these difficulties, there are many ongoing and planned studies of cosmic shear measurement with the Hyper Suprime-Cam on Subaru (http://www.naoj.org/Projects/HSC/HSCProject.html),  EUCLID (Euclid http://sci.esa.int/euclid), and LSST (http://www.lsst.org), with the hope that an accurate measurement method will be available by the time of the observations. 
Many investigators have been developing such methods (Kaiser et al 1995, Bernstein \& Jarvis 2002; Refregier 2003; Kuijken et al. 2006; Miller et al. 2007; Kitching et al. 2008; Melchior 2011), 
and some of them are being tested using simulated data (Heymans et al 2006, Massey et al 2007, Bridle et al 2010 and Kitching et al 2012).
We have also  developed a new shear analysis method (E-HOLICs, Okura and Futamase 2011, Okura and Futamase 2012, Okura and Futamase 2013), and have partially succeeded in avoiding the systematic error 
arising from inaccurate shape measurements by adopting an elliptical weight function.  
The other important cause of the systematic errors in the moment method comes from an unjustifiable approximation adopted in the point spread function (PSF) correction.  
Recently, we developed a new method for PSF correction referred to as the ellipticity of re-smeared artificial images (ERA) method.  
It makes use of the artificial image constructed by re-smearing the lensed image, and the artificial image   has the same ellipticity with the lensed image (ERA: Okura and Futamase 2014).
We have shown that the method  avoids, in principle, the systematic error associated with the PSF effect. 

In this paper, we propose a new weak lensing measurement scheme that uses a newly defined ellipticity in the ERA method. 
The ellipticity we use in this technique is defined by the 0th order moment of the shape and  is therefore the ellipticity  associated with  
a region near the centre of the image, rather than the usual 
ellipticity defined by the 2nd order moment of the shape. We expect that this ellipticity has a higher signal-to-noise ratio  (S/N)
compared with the usual ellipticity because the central region is brighter than the surrounding region in general.  
We confirmed this expectation using actual data from the galaxy cluster Abell 1689 obtained from  Subaru.

This paper is organized as follows.
In Sect. 2, we explain the notation and some basic concepts used in this paper for the sake of completeness. 
We then define the new ellipticity using the 0th order moment of the image and its relationship to the gravitational shear as well as to 
the usual ellipticity in Section 3. 
In Section 4, we explain the merit of using the ellipticity from the point of view of the S/N. 
In the actual measurement of the 0th order moment there is some difficulty. We  explain the difficulty and its solution 
in Section 5.  
Then we show the results of measurement of the 0th-ellipticity using  data from Abell1689 in Section 6. 
In Section 7, we explain  PSF correction for the 0th-ellipticity and give the results of  tests of PSF correction for the ellipticity adopted in this paper. 
Finally, Section 8 is devoted to the conclusion and some comments. 
%----------------------------------------------------------------------------------------------------
%----------------------------------------------------------------------------------------------------
%----------------------------------------------------------------------------------------------------
\section{Notation and basic concepts}
In this section we briefly explain the notations, basic concepts, and definitions used in this paper for 0th order moments and the ellipticity defined by these moments (0th-ellipticity). 
The 0th-ellipticity is defined by image moments and we use the idea of zero plane.
The notation and definition are the same as in  Okura \& Futamase 2014, so further details can be seen in that paper.

The concept of the zero plane and the zero image that we propose assumes that the intrinsic image with the reduced shear ($\mbg^I$) is the result of an imaginary distortion of the corresponding zero image with 0 ellipticity. The zero plane is an imaginary plane where zero images are located. 
Thus we have three kinds of planes: zero, source, and image planes.  
We use the complex coordinate $\bbe=\beta_1 +i\beta_2$ in the zero plane, $\bbe^s=\beta^s_1 +i\beta^s_2$ in the source plane, and $\bth=\theta_1 +i\theta_2$ in the image plane, and the relations between the coordinates are obtained as
\begin{eqnarray}
\label{eq:CORs}
\bbe&=&\bbe^s-\mbg^I\bbe^{s*}\\
\bbe^s&=&\bth-\mbg^L\bth^*
,\end{eqnarray}
where $\mbg^L$ is the lensing shear.
We set the origin of the coordinates at the centroid of the zero image and the intrinsic image, 
where the centroid is defined by the condition that the dipole moments of images 
vanish. The detailed definition can be seen in Okura and Futamase 2014.
The combined shear, which is a distortion from the zero plane to the image plane, is written as $\mbg$, 
which is given by the lensing shear $\mbg^L$ and the intrinsic shear as follows:
\begin{eqnarray}
\label{eq:convg}
\mbg \equiv \frac{\mbg^I+\mbg^L}{1+\mbg^I\mbg^{L*}}.
\end{eqnarray}
The intrinsic reduced shear $\mbg^I$ is regarded as a value from intrinsic ellipticity, so it has random orientation. This means the average value of the intrinsic shear tends to 0, so we can estimate lensing shear by averaging Eq. \ref{eq:convg}.

The complex moments of the measured image are denoted as $\cZ^N_M$ and measured as
\begin{eqnarray}
\label{eq:CMOM}
\cZ^N_M(I,\bep_W)&\equiv&\int d^2\theta \bth^N_M I(\bth) W(\bth,\bep_W)\\
\label{eq:CMOM2}
\bth^N_M&\equiv&\bth^{\frac{N+M}{2}}\bth^{*\frac{N-M}{2}},
\end{eqnarray}
where W is a weight function, which is a function of displacement from the centroid $\bth$ and ellipticity $\bep_W$. 
The determination of the ellipticity is explained in the next section and the Appendix,
 and the subscript N is the order of moments and M indicates the spin number. 
For example:
 
\begin{eqnarray}
\cZ^2_0(I,\bep_W)&\equiv&\int d^2\theta \bth\bth^* I(\bth) W(\bth,\bep_W)\nonumber\\
&=&\int d^2\theta \lr{\theta_1^2+\theta_2^2} I(\bth) W(\bth,\bep_W)\\
\cZ^2_2(I,\bep_W)&\equiv&\int d^2\theta \bth^2 I(\bth) W(\bth,\bep_W)\nonumber\\
&=&\int d^2\theta \lr{\theta_1^2-\theta_2^2+2i\theta_1\theta_2} I(\bth) W(\bth,\bep_W)
.\end{eqnarray}

Thus the usual ellipticity is defined from $\cZ^2_2$ and $\cZ^2_0$.  
The weight function has an arbitrary profile and size, but it should be chosen to reduce the effect from pixel noise.
In the simulation test in this paper, we use an elliptical Gaussian for the weight function. %----------------------------------------------------------------------------------------------------
%----------------------------------------------------------------------------------------------------
%----------------------------------------------------------------------------------------------------
\section{Ellipticity of the 0th order moment}
In this section, we define the new ellipticity and its relationship to the usual ellipticity and weak gravitational lensing shear without PSF effect. Weak lensing shear analysis by 0th-ellipticity with PSF correction is explained in section 7.

%----------------------------------------------------------------------------------------------------
\subsection{Definition}
As mentioned in the Introduction, the new ellipticity is defined from the 0th order moments. 
However, ellipticity must have a property of spin 2, and it seems impossible to derive a spin 2 
quantity from monopole moments $\cZ^0_0$. In fact, this is not so.
The spin 2 quantity can be constructed as follows. 
First, we define the 0th order moment that has spin 2 property as
\begin{eqnarray}
\label{eq:Z02}
\cZ^0_2(I,\bep_W)&\equiv&\int d^2\theta\frac{\bth^2_2}{\bth^2_0}I(\bth) W(\bth,\bep_W)\nonumber\\
&=&\int d^2\theta\lr{\cos (2\phi_\theta)+i\sin (2\phi_\theta)}I(\bth) W(\bth,\bep_W),
\end{eqnarray}
where $\phi_\theta$ is the position angle at $\bth$. 
It has spin 2 nature and thus the ellipticities are defined by normalizing with monopole moments($\cZ^0_0$) to have non-dimension,
\begin{eqnarray}
\label{eq:E0th}
\mbeZ&\equiv&\lrs{\frac{\cZ^0_2}{\cZ^0_0}}_{(I,\bep_W)}\\
\label{eq:EP0th}
\bepZ&\equiv& 2\mbeZ/(1+|\mbeZ|^2),
\end{eqnarray}
where $\mbeZ$ is the ellipticity measured from 0th order moments and
$\bepZ$ is another ellipticity used for the ellipticity of the weight function $\bep_W=\bepZ$ in Eq. \ref{eq:Z02}, 
and the reason for using this ellipticity for weight function is explained in Appendix \ref{AP:0thE}.
The values of $\mbeZ$ and $\bepZ$ are different,  but they provide the same information regarding the ellipticity so they can be transformed into each other.

To distinguish the ellipticities defined by 0th order moments from the ellipticities defined by quadrupole moments, we denote ellipticities defined by quadrupole moments as
\begin{eqnarray}
\label{eq:EP2nd}
\bep_{2nd}&\equiv&\lrs{\frac{\cZ^2_2}{\cZ^2_0}}_{\lr{I,\bep_W}}\\
\label{eq:E2nd}
\mbe_{2nd}&=&\frac{\bep_{2nd}}{|\bep_{2nd}|^2}\lr{1-\sqrt{1-|\bep_{2nd}|^2}}
,\end{eqnarray}
where the ellipticity of the weight function in eq.\ref {eq:EP2nd} is used and the ellipticity is defined by quadrupole moments, so $\bep_W=\bep_{2nd}$.

We refer to $\mbeZ$ and $\bepZ$ as the "0th-ellipticity" and $\mbe_{2nd}$ and $\bep_{2nd}$ as the "2nd-ellipticity".
If the profile of the measured image is simple, for example an elliptical Gaussian, the 
0th- and 2nd-ellipticities have the same value ($\mbeZ=\mbe_{2nd}$ and $\bep_{0th}=\bep_{2nd}$),
but because a real image has a complex form, these ellipticities usually have different values.

%----------------------------------------------------------------------------------------------------
\subsection{0th-ellipticity and reduced shear}
Here we present the relationship between the 0th-ellipticity and the reduced shear without the PSF effect.
We assume $\ILnsdt$ to be an image distorted by lensing reduced shear $\mbg$ from the zero image $\IZt$.
From the definitions of the 0th-ellipticity, the 0th order moments satisfy the following relationship:
\begin{eqnarray}
\label{eq:evi0thimage}
\cZ^0_2(\ILnsd,\bep_W)-\mbeZ\cZ^0_0(\ILnsd,\bep_W)=0.
\end{eqnarray}
By transforming eq.\ref{eq:evi0thimage} to the zero plane we obtain
\begin{eqnarray}
(\mbg-\mbeZ)\int d^2\beta\frac{\bbe^2_0}{\bbe^2_0+\mbg\bbe^{2*}_2}\IZt W^0(\bbe,0)=0&&\hspace{1pt}g<1\\
\frac{1-\mbg^*\mbeZ}{\mbg^*}\int d^2\beta\frac{\bbe^2_0}{\bbe^2_0+\bbe^{2*}_2/\mbg^*}\IZt W^0(\bbe,0)=0&&\hspace{1pt}g>1 
.\end{eqnarray}
Thus, the relationship between the 0th-ellipticity and the reduced shear is obtained as follows:
\begin{eqnarray}
\mbeZ=&\mbg&\hspace{50pt}g<1
\\
\mbeZ=&\frac{1}{\mbg^*}&\hspace{50pt}g>1, 
\\
\bepZ=&\bde&\equiv{2\mbg}/\lr{1+g^2}
.\end{eqnarray}

Detailed calculations are shown in  Appendix \ref{AP:0thE}.
This result shows that the reduced shear can be obtained from the 0th-ellipticity.
At the same time, the reduced shear can also be obtained from the 2nd-ellipticity.
Since the 2nd-ellipticity is induced by the same lensing effect, more accurate weak lensing 
analysis will be achieved by combining these two ellipticities simultaneously. 

%----------------------------------------------------------------------------------------------------
%----------------------------------------------------------------------------------------------------
%----------------------------------------------------------------------------------------------------
\section{Effective signal-to-noise ratio}
In this section, we show  improvement of the S/N in the measurement of 
the 0th-ellipticity compared with the 2nd-ellipticity.

The regions for measuring the monopole (quadrupole) moment and 0th (2nd)-ellipticity are the same.
 Therefore the S/N of the 0th (2nd)-ellipticity $\rSN_{0th}(\rSN_{2nd})$ is the same 
as that of the monopole(quadrupole) moments.

Usually the S/N is defined as
\begin{eqnarray}
\label{eq:SN}
\rSN=\frac{\int d^2\theta I(\bth) W(\bth,\bep_W)}{\sigma_N\sqrt{\int d^2\theta W^2(\bth,\bep_W)}},
\end{eqnarray}
where $\sigma_N$ is the standard deviation of pixel noise,
and therefore this is also the S/N of the monopole moment. Thus we obtain $\rSN_{0th}=\rSN$.
However, the quadrupole moment has a different S/N from that of the monopole moment,
because the region used to measure the monopole and quadrupole moments are different.
Therefore, the quadrupole moment and 2nd-ellipticity have different signal and noise from $\rSN_{0th}$, 
and is defined as 

\begin{eqnarray}
\rSN_{2nd}=\frac{\int d^2\theta I(\bth) \bth^2_0W(\bth,\bep_W)}{\sqrt{\int d^2\theta \bth^4_0 W^2(\bth,\bep_W)}}.
\end{eqnarray}

Now we consider an image which has a Gaussian brightness distribution
to analytically compare  these S/Ns. 
The ratio of the S/Ns for a Gaussian image with the same weight function is given by 

\begin{eqnarray}
\frac{\rSN_{2nd}}{\rSN_{0th}}=\frac{1}{\sqrt{2+\epsilon_W^2}}.
\end{eqnarray}

This result means that the S/N of the quadrupole moment is lower than that of the monopole moment 
on the order of $\sqrt{2}\sim\sqrt{3}$
because the monopole moment is measured in the central (brighter) region. 
Therefore, it is expected that the reduced shear can be measured by using the 0th-ellipticity with a higher S/N compared to using the 2nd-ellipticity.
We measured the S/N  with real data in Section \ref{sec:REALDATA}.

%----------------------------------------------------------------------------------------------------
%----------------------------------------------------------------------------------------------------
%----------------------------------------------------------------------------------------------------
\section{Integration problem}
In this section, we present a problem and a correction in measuring moments by integration 
of the 0th-ellipticity, which is distinct from the sampling or pixelization problem considered in typical image analysis.

Moments are defined by integration of the image count and some functions of distance from the 
centroid, e.g. eq. \ref{eq:CMOM}.
However, in the  analysis of actual data, we do not use  integration to measure the moments,
because the image is observed by CCD pixels. 
Thus, instead of the integration, 
we use a sum of sampling counted CCD pixels, and eq. \ref{eq:Z02} becomes
\begin{eqnarray}
\label{eq:Z02S}
\cZ^0_2(I,\bep_W)
&=&\sum_x\sum_y \frac{{\bth_{xy}}^2_2}{{\bth_{xy}}^2_0}I(\bth_{xy}) W(\bth_{xy},\bep_W)
\nonumber\\
&=&\sum_x\sum_y \lr{\cos(2\phi_{\theta xy}+i\sin(2\phi_{\theta xy})}I(\bth_{xy}) W(\bth_{ixy},\bep_W)
,\end{eqnarray}
where $\bth_{xy}=\theta_x+i\theta_y$ is the displacement of position in the   bottom left corner of each pixel from the centroid.
However, ${{\bth_{xy}}^2_2}/{{\bth_{xy}}^2_0}$ changes violently in a pixel, 
especially if the CCD pixels are near the centroid, 
then  the sampling count in eq. \ref{eq:Z02S} makes an incorrect count for the moment.
In measuring the quadrupole moments, this effect is sufficiently small, 
because ${\bth_{xy}}^2_2$ takes almost 0 at pixels near the centroid.
To correct this effect, we use integration only for ${{\bth_{xy}}^2_2}/{{\bth_{xy}}^2_0}$ and eq. \ref{eq:Z02S} is redefined as
\begin{eqnarray}
\cZ^0_2(I,\bep_W)
&=&\sum_x\sum_y I(\bth_{xy}) W(\bth_{xy},\bep_W)
\int_{\theta_x}^{\theta_x+1}\hspace{-15pt}d\theta'_x
\int_{\theta_y}^{\theta_y+1}\hspace{-15pt}d\theta'_y \hspace{5pt} \frac{{\bth'}^2_2}{{\bth'}^2_0}
\\&=&\sum_x\sum_y I(\bth_{xy}) W(\bth_{xy},\bep_W)
\int_{\theta_x}^{\theta_x+1}\hspace{-15pt}d\theta'_x
\int_{\theta_y}^{\theta_y+1}\hspace{-15pt}d\theta'_y 
\nonumber\\&&
\hspace{75pt}\times \frac{{{\theta'}_x}^2-{{\theta'}_y}^2+2i{{\theta'}_x}{{\theta'}_y}}{{{\theta'}_x}^2+{{\theta'}_y}^2}
\\&\equiv&\sum_x\sum_y I(\bth_{xy}) W(\bth_{xy},\bep_W)
\lr{C_R(\bth_{xy})+iC_I(\bth_{xy})}.
\end{eqnarray}
This integration can be calculated analytically as 
\begin{eqnarray}
\int_{\theta_x}^{\theta_x+1}\hspace{-15pt}d\theta'_x
\int_{\theta_y}^{\theta_y+1}\hspace{-15pt}d\theta'_y \hspace{5pt} \frac{{{\theta'}_x}^2-{{\theta'}_y}^2}{{{\theta'}_x}^2+{{\theta'}_y}^2}
\hspace{-75pt}&&\nonumber\\
&=&\Biggl[\left[\lr{{\theta'}_x^2+{\theta'}_y^2}\tan^{-1}\lr{\frac{{\theta'}_y}{{\theta'}_x}}\right]^{\theta_x+1}_{\theta_x}\Biggr]^{\theta_y+1}_{\theta_y}
\hspace{20pt}{\theta'}_x\neq0\\
&\equiv&\Bigl[\left[F_A(\theta_x,\theta_y)\right]^{\theta_x+1}_{\theta_x}\Bigr]^{\theta_y+1}_{\theta_y}
\hspace{70pt}{\theta'}_x\neq0\\
\int_{\theta_x}^{\theta_x+1}\hspace{-15pt}d\theta'_x
\int_{\theta_y}^{\theta_y+1}\hspace{-15pt}d\theta'_y \hspace{5pt} \frac{{{\theta'}_x}^2-{{\theta'}_y}^2}{{{\theta'}_x}^2+{{\theta'}_y}^2}
\hspace{-75pt}&&\nonumber\\
&=&-\Biggl[\left[\lr{{\theta'}_x^2+{\theta'}_y^2}\tan^{-1}\lr{\frac{{\theta'}_x}{{\theta'}_y}}\right]^{\theta_x+1}_{\theta_x}\Biggr]^{\theta_y+1}_{\theta_y}
\hspace{15pt}{\theta'}_y\neq0\\
&=&\Bigl[\left[F_B(\theta_x,\theta_y)\right]^{\theta_x+1}_{\theta_x}\Bigr]^{\theta_y+1}_{\theta_y}
\hspace{70pt}{\theta'}_y\neq0\\
\int_{\theta_x}^{\theta_x+1}\hspace{-15pt}d\theta'_x
\int_{\theta_y}^{\theta_y+1}\hspace{-15pt}d\theta'_y \hspace{5pt} \frac{2{{\theta'}_x}{{\theta'}_y}}{{{\theta'}_x}^2+{{\theta'}_y}^2}
\hspace{-75pt}&&\nonumber\\
&=&\frac{1}{2}
\Bigl[\left[\lr{{\theta'}_x^2+{\theta'}_y^2}\log\lr{{{\theta'}_x}^2+{{\theta'}_y}^2}\right]^{\theta_x+1}_{\theta_x}\Bigr]^{\theta_y+1}_{\theta_y}\\
&\equiv&\frac{1}{2}\Bigl[\left[F_C(\theta_x,\theta_y)\right]^{\theta_x+1}_{\theta_x}\Bigr]^{\theta_y+1}_{\theta_y};
\end{eqnarray}
therefore, we can obtain $C_R$ and $C_I$ analytically as
\begin{eqnarray}
C_R(\bth_{xy})
\hspace{-25pt}&&\nonumber\\
&=&\Bigl[\left[F_A(\theta_x,\theta_y)\right]^{\theta_x+1}_{\theta_x}\Bigr]^{\theta_y+1}_{\theta_y}
\hspace{15pt}{\rm if}\hspace{5pt}|\theta_x+0.5|>|\theta_y+0.5|,\\
C_R(\bth_{xy})
\hspace{-25pt}&&\nonumber\\
&=&\Bigl[\left[F_B(\theta_x,\theta_y)\right]^{\theta_x+1}_{\theta_x}\Bigr]^{\theta_y+1}_{\theta_y}
\hspace{15pt}{\rm if}\hspace{5pt}|\theta_y+0.5|>|\theta_x+0.5|
\\
\label{eq:center}
C_R(\bth_{xy})
\hspace{-25pt}&&\nonumber\\
&=&\frac12\Biggl(
\Bigl[\left[F_A(\theta_x,\theta_y)\right]^{\theta_x+1}_{\theta_x}\Bigr]^{\theta_y+1}_{\theta_y}+
\Bigl[\left[F_B(\theta_x,\theta_y)\right]^{\theta_x+1}_{\theta_x}\Bigr]^{\theta_y+1}_{\theta_y}
\nonumber\\&&
+\pi\lr{
  \frac{\lr{\theta_x+1}^3}{|\theta_x+1|}
-\frac{\lr{\theta_x    }^3}{|\theta_x     |}
-\frac{\lr{\theta_y+1}^3}{|\theta_y+1|}
+\frac{\lr{\theta_y   }^3}{|\theta_y     |}
}\Biggr)
\nonumber\\&&
\hspace{50pt}{\rm if}\hspace{5pt}(-1<\theta_x<0) \hspace{5pt}{\rm and}\hspace{5pt} (-1<\theta_y<0)
\\
C_I(\bth_{xy})
\hspace{-25pt}&&\nonumber\\
&=&\frac{1}{2}\Bigl[\left[F_C(\theta_x,\theta_y)\right]^{\theta_x+1}_{\theta_x}\Bigr]^{\theta_y+1}_{\theta_y}
,\end{eqnarray}
where eq. \ref{eq:center} is a formula about integration in the pixel, which is the centre of image.
This formula can be derived as
\begin{eqnarray}
C_R(\bth_{xy})
\hspace{-25pt}&&\nonumber\\
&=&\frac12\lim_{\delta\rightarrow 0}\Biggl(
\Bigl[\left[F_A(\theta_x,\theta_y)\right]^{-\delta}_{\theta_x}+
\left[F_A(\theta_x,\theta_y)\right]^{\theta_x+1}_{\delta}\Bigr]^{\theta_y+1}_{\theta_y}
\nonumber\\&&\hspace{50pt}+
\Bigl[\left[F_B(\theta_x,\theta_y)\right]^{-\delta}_{\theta_y}+
\left[F_B(\theta_x,\theta_y)\right]^{\theta_y+1}_{\delta}\Bigr]^{\theta_x+1}_{\theta_x}
\Biggr)\nonumber\\&=&
\frac12\Biggl(
\Bigl[\left[F_A(\theta_x,\theta_y)+F_B(\theta_x,\theta_y)\right]^{\theta_y+1}_{\theta_y}\Bigr]^{\theta_x+1}_{\theta_x}
\nonumber\\&&
+\lim_{\delta\rightarrow 0}\Biggl(-
\lr{\theta_y+1}^2\arctan\lr{ \frac{\theta_y+1}{\delta}}+
\lr{\theta_y+1}^2\arctan\lr{-\frac{\theta_y+1}{\delta}}
\nonumber\\&&\hspace{30pt}+
\lr{\theta_y}^2\arctan\lr{ \frac{\theta_y}{\delta}}-
\lr{\theta_y}^2\arctan\lr{-\frac{\theta_y}{\delta}}
\nonumber\\&&\hspace{30pt}+
\lr{\theta_x+1}^2\arctan\lr{ \frac{\theta_x+1}{\delta}}-
\lr{\theta_x+1}^2\arctan\lr{-\frac{\theta_x+1}{\delta}}
\nonumber\\&&\hspace{30pt}-
\lr{\theta_x}^2\arctan\lr{ \frac{\theta_x}{\delta}}+
\lr{\theta_x}^2\arctan\lr{-\frac{\theta_x}{\delta}}
\Biggr)\Biggr)
\nonumber\\&=&
\frac12\Biggl(
\Bigl[\left[F_A(\theta_x,\theta_y)+F_B(\theta_x,\theta_y)\right]^{\theta_y+1}_{\theta_y}\Bigr]^{\theta_x+1}_{\theta_x}
\nonumber\\&&\hspace{50pt}+
\pi\lr{
\frac{\lr{\theta_x+1}^3}{|\theta_x+1|}-\frac{\lr{\theta_x}^3}{|\theta_x|}
-\frac{\lr{\theta_y+1}^3}{|\theta_y+1|}+\frac{\lr{\theta_y}^3}{|\theta_y|}
}
\Biggr)
.\end{eqnarray}

%----------------------------------------------------------------------------------------------------
%----------------------------------------------------------------------------------------------------
%----------------------------------------------------------------------------------------------------
\section{Statistics obtained from real data}
\label{sec:REALDATA}
In  this section, we show the statistics of the ellipticities and 
signal-to-noise ratios obtained using data from Abell1689.

\subsection{Data}
 We used real data from the field of Abell 1689, which was observed by the Subaru telescope and Suprime-Cam.
Objects were detected by Sextractor and we selected objects that are larger than  stars.
The region of the data is about 0.245(deg$^2$) and the number of the selected background objects is 22302 ($\sim$ 25 /arcmin$^2$).

\subsection{Statistics}
\begin{figure*}[htbp]
\centering
\resizebox{\hsize}{!}{\includegraphics{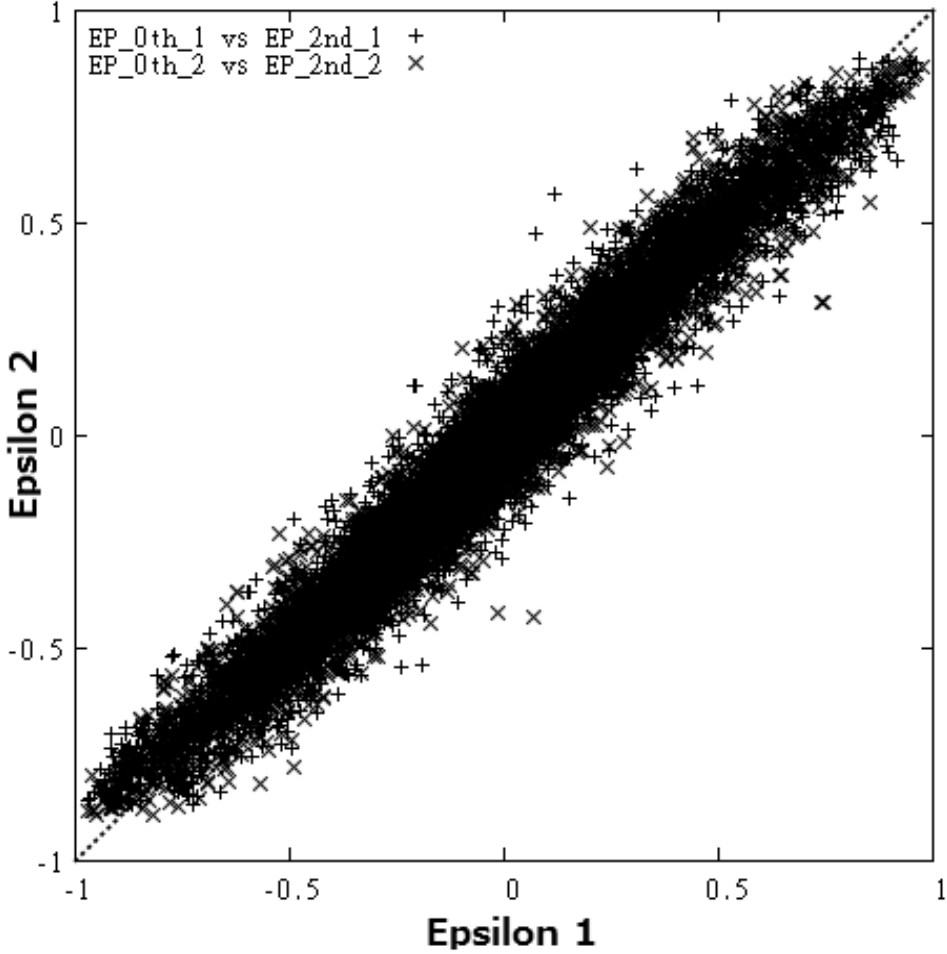}}
\caption{
\label{fig:WGLSA_SHAPE_EP}
The correlation between 0th- and 2nd-epsilon values measured from real data
}
\end{figure*}

We now present the statistics of the ellipticities and the signal-to-noise ratios 
for the selected objects.
Figure \ref{fig:WGLSA_SHAPE_EP}  shows the correlation between the 0th- and 2nd-ellipticites,
where the  0th-ellipticities are plotted on the horizontal axis and the  2nd-ellipticities are on the vertical axis. 
A cut-off for objects that have $|\bep_{2nd}|>0.9$ was used.
The figure shows a high correlation,
and thus we can equally use the 0th-ellipticity as a tool for weak shear
measurement in the analysis of real data.

\begin{figure*}[htbp]
\centering
\resizebox{\hsize}{!}{\includegraphics{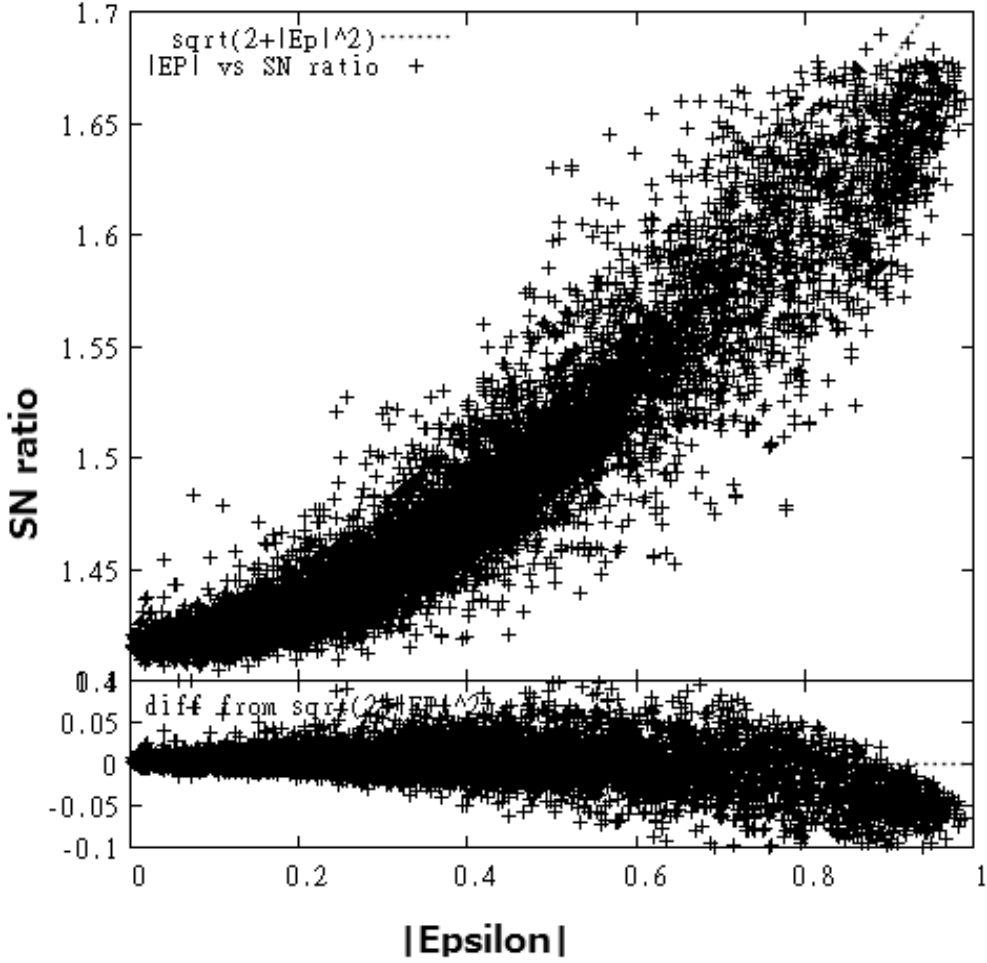}}
\caption{
\label{fig:WGLSA_SHAPE_SNR}
Epsilon vs S/N measured from real data
}
\end{figure*}

\begin{figure*}[htbp]
\centering
\resizebox{\hsize}{!}{\includegraphics{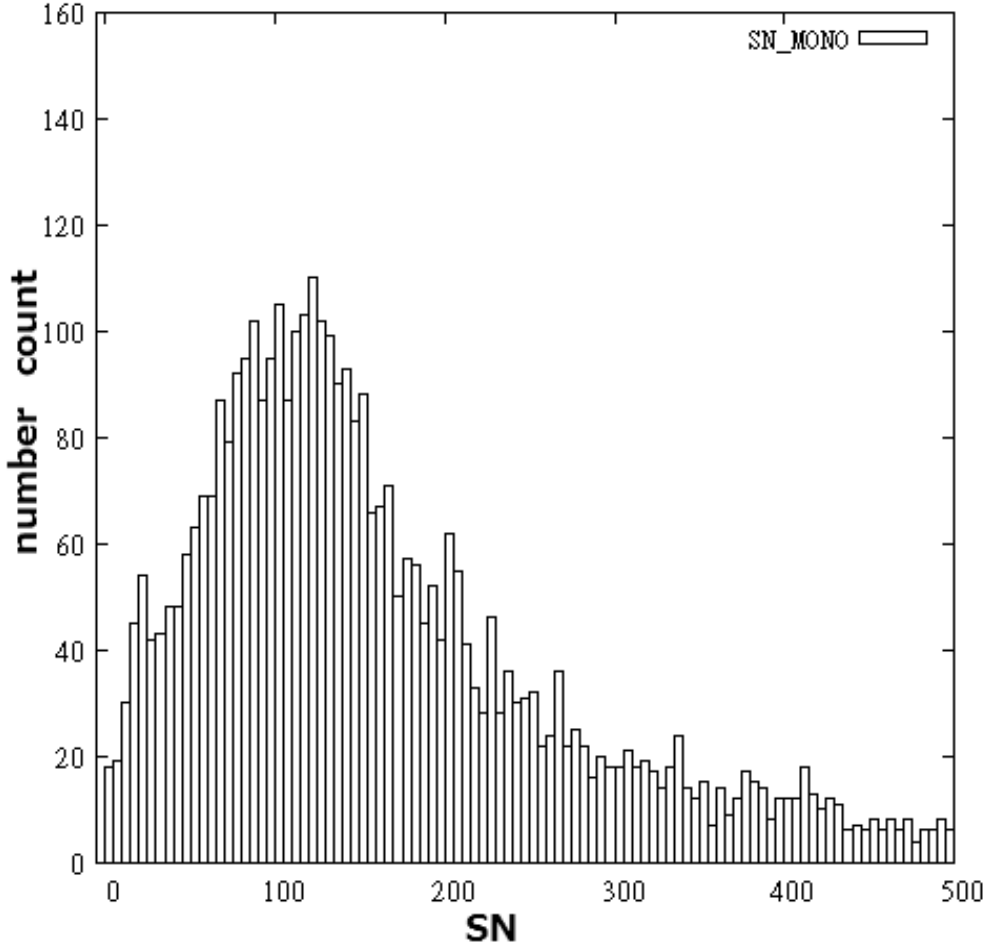}}
\caption{
\label{fig:WGLSA_SHAPE_SN_MONO}
The distribution of the normalized number count  of S/N monopole
}
\end{figure*}

\begin{figure*}[htbp]
\centering
\resizebox{\hsize}{!}{\includegraphics{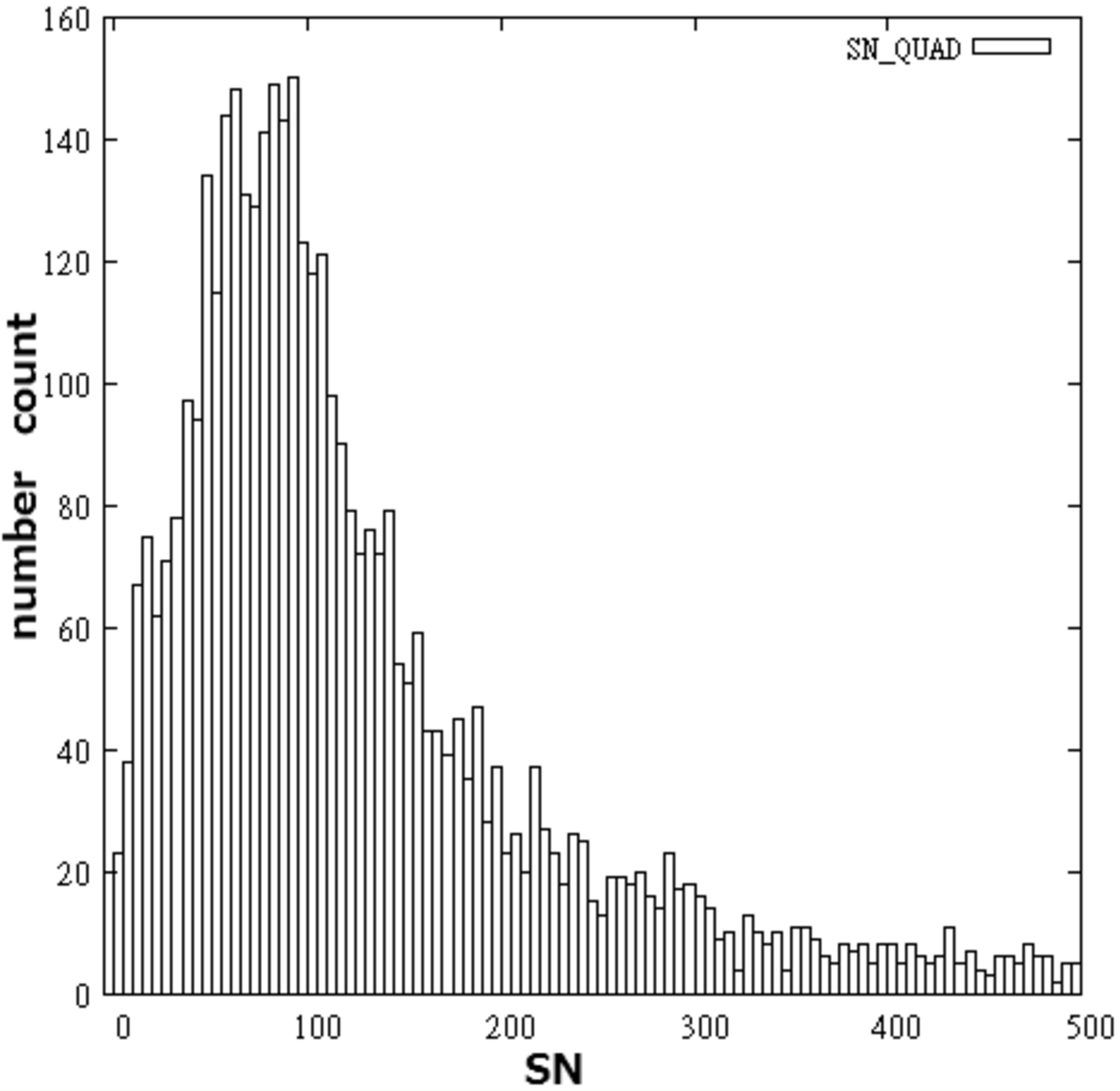}}
\caption{
\label{fig:WGLSA_SHAPE_SN_QUAD}
The distribution of the normalized number count of S/N quadrupole
}
\end{figure*}

The upper part of Figure \ref{fig:WGLSA_SHAPE_SNR} shows the relationship between 
the ellipticity and the ratio between  $\rSN_{0th}$ and $\rSN_{2nd}$,
where  $|\bep_{0th}|$ is plotted on the horizontal axis and  the ratio $\rSN_{0th}/\rSN_{2nd}$ is on the vertical axis.
The dashed line in the figure is $\sqrt{2+|\bep_{0th}|^2}$, which 
is the predicted function for an elliptical Gaussian image.
The bottom part of Figure \ref{fig:WGLSA_SHAPE_SNR} shows the difference in the S/N  from $\sqrt{2+|\bepZ|^2}$.
These figures show that the prediction is a good approximation for real objects.
The values of $\rSN_{0th}$ for real objects are $\sqrt{2}\sim\sqrt{3}(\sim 1.5)$ times larger than $\rSN_{2nd}$. Figures \ref{fig:WGLSA_SHAPE_SN_MONO} and  \ref{fig:WGLSA_SHAPE_SN_QUAD}
show the  distributions of the number count for each of the S/N bins.
It can be seen that the number count of $\rSN_{0th}$ is distributed in higher S/N bins than that of $\rSN_{2nd}$,
which is consistent with Figure \ref{fig:WGLSA_SHAPE_SNR}.
Based on the above measurements,  the 0th-ellipticity has a 1.5 times larger S/N than that of the 2nd-ellipticity,
which means that the lower limit of the S/N can be set to a 1.5 times smaller
value for the 0th-ellipticity with the same precision of pixel noise.
Therefore, it is expected that  
the available number of background objects will increase, and as a result,  the statistical error in
the  weak lensing shear measurement scheme  is lower than with the
previous method. A more detailed and qualitative study of  error
estimation will be given in a forthcoming paper.

%----------------------------------------------------------------------------------------------------
%----------------------------------------------------------------------------------------------------
%----------------------------------------------------------------------------------------------------
%----------------------------------------------------------------------------------------------------
%----------------------------------------------------------------------------------------------------
%----------------------------------------------------------------------------------------------------
%----------------------------------------------------------------------------------------------------
%----------------------------------------------------------------------------------------------------

%----------------------------------------------------------------------------------------------------
%----------------------------------------------------------------------------------------------------
%----------------------------------------------------------------------------------------------------
\section{PSF correction}
In the previous section we showed the relationship between the 0th-ellipticity and the reduced shear. 
The 0th-ellipticity is simply the reduced shear if the reduced shear is 
less than 1 and if there is no PSF effect. We will now account for the PSF effect. 
In this paper we use the ERA method which is described in Okura and Futamase 2014 in detail, and we briefly outline this method.   

%----------------------------------------------------------------------------------------------------
\subsection{ERA method}
We first write the image, which was smeared by PSF from the lensed image, as $\ISmdt$. This image can be described by convolution with the lensed image and PSF $P(\bth)$ as
\begin{eqnarray}
\hISmdk = \hILnsdk \hat P(\mbk),
\end{eqnarray}
where the ``hat'' symbol indicates  that a Fourier transform has been performed on the function.
We denote the ellipticity of the observed image $\ISmd$ as $\bep^{S}_{0th}$,
and the ellipticity of the lensed image $\ILnsd$ as $\bep^{L}_{0th}$.
One possible way to obtain $\bep^{L}_{0th}$ is by deconvolution such as
\begin{eqnarray}
\label{eq:DEC}
\hILnsdk=\frac{\hISmdk}{\hat P(\mbk)}=\frac{\hISmdk\hat P^*(\mbk)}{|\hat P(\mbk)|^2}.
\end{eqnarray}
We need to introduce the deconvolution constant $C^{Dec}$ to avoid dividing by zero, so that
\begin{eqnarray}
\label{eq:CDEC}
\hILnsdk=\frac{\hISmdk\hat P^*(\mbk)}{|\hat P(\mbk)|^2+C^{Dec}}.
\end{eqnarray}
Especially in real analysis, a large deconvolution constant is needed owing to pixel noise,
which makes a different ellipticity and introduces a systematic error;
some samples of the errors from large $C^{Dec}$ can be seen in Table 1, but the table shows only a tendency introduced by using large $C^{Dec}$, and the detailed values of $C^{Dec}$ do not have any meaning, because it depends on total count of images and so on.

The idea of PSF correction for the 0th-ellipticity is to construct an artificial image $\IESmd$, which has $\bep^{L}_{0th}$ by re-smearing the lensed image using an artificial PSF $P^E(\bth,\bep^L_{0th})$ 
with ellipticity $\bep^{L}_{0th}$ (but an arbitrary profile and size) in such a way that 
the following relation is satisfied: 
\begin{eqnarray}
\label{eq:ERA}
\hIESmdk = \hILnsdk \hat P^E(\mbk) =\hISmdk \frac{ \hat P^E(\mbk)}{\hat P(\mbk)}.
\end{eqnarray}
Since $\IESmd$ also has the 
ellipticity $\bep^{L}_{0th}$ by definition, the reduced shear can be obtained by measuring $\bep^{ES}_{0th}$.
In the previous ERA paper, we suggested two possible methods for solving eq. \ref{eq:ERA}. 
One is to use the deconvolution constant, which we refer to as  Method A.
Another is  re-smearing  the deconvolved image, which  we call Method B. 
In Method B, the artificial PSF is divided as $\hat P^E(\mbk)=\hat P(\mbk) \Delta\hat P(\mbk)$;
then eq. \ref{eq:ERA} becomes 
\begin{eqnarray}
\label{eq:ERAB}
\hIESmdk =\hISmdk \Delta\hat P(\mbk).
\end{eqnarray}

Both methods require iterations until $\IESmd$ and $P^E$ have the same ellipticity.
We refer to the ERA method of PSF correction applied to the 0th-ellipticity as the 0-ERA method. 
More details regarding the ERA method can be found in the previous paper. 

\begin{table}
\begin{tabular}{|c|c|c|c|c|}\hline
$C^{Dec}$ & $10^{-7}$  & $10^{-6}$ &$10^{-5}$ & $10^{-4}$
\\\hline
$\epsilon$& 0.299 & 0.298 & 0.295 & 0.286
\\\hline\hline
$C^{Dec}$  & $10^{-3}$ & $10^{-2}$ & $10^{-1}$ & $10^{0}$
\\\hline
$\epsilon$& 0.279 & 0.265 & 0.224 & 0.168
\\\hline
\end{tabular}
\caption{The tendency of incorrect deconvolution by using relatively higher values for $C^{Dec}$.}
\end{table}

%----------------------------------------------------------------------------------------------------
\subsection{Simulation test}
We tested a precision PSF correction for the 0th-ellipticity by using a simple simulated image.
In real weak lensing analysis, there are many effects that make errors in shear estimation, e.g. PSF effect, pixel noise effect, pixelization, etc.  Therefore, we need to study  them for precise shear estimation, but many of them are too complex  to study at the same time. So in this paper, we consider only simple PSF effects because the ERA method does not use  approximations in PSF correction; then, if the results of this simulation have no systematic errors it shows that 0th-ellipticity can be used in the  ERA method in the  same way as 2nd-ellipticity.
The parameters for this simulation are the same as used in Okura and Futamase 2014 which initially described the ERA method. 

\subsection{Simulation parameters}
We used Gaussian and S\'{e}rsic galaxy images and a Gaussian PSF image, where
the ellipticity of the simulated image $\bep_{0th}$ is (0.3, 0.0);
the ellipticities of the PSF are [(0.0, 0.0), (0.1, 0.1), (0.3, 0.0), (-0.6, -0.6)]; and
the size of the PSF is 0.5, 1.0, and 1.5 times as large as the galaxy. 
A large size image is used to enable us to neglect pixelization.
We then tested the following six  cases.

\begin{itemize}
\item 1. Deconvolution: standard deconvolution with a deconvolution constant
small enough to neglect error from the deconvolution constant  discussed in 
section 3.3.
This is not a realistic analysis owing to the pixel noise effect.
\item 2. Method A: re-smearing the deconvolved image where the size of $P^E$ is the same as the PSF $R^P$.
\item 3. Method A2: re-smearing the deconvolved image where the size of $P^E$ is 2 times as large as the PSF $R^P$.
\item 4. Method A3: re-smearing the deconvolved image where the size of $P^E$ is 3 times as large as the PSF $R^P$.
\item 5. Method B1: re-smearing the smeared image where size of $\Delta P$ is the same as the PSF $R^P$. 
\item 6. Method B2: re-smearing the smeared image where the size of $\Delta P$ is the same as the size of $\ISmd$.
\end{itemize}

This simulation  is the same as that described in the previous paper where  the ERA method was described. More detailed information is available in that paper.

\begin{figure*}[htbp]
\centering
\resizebox{\hsize}{!}{\includegraphics{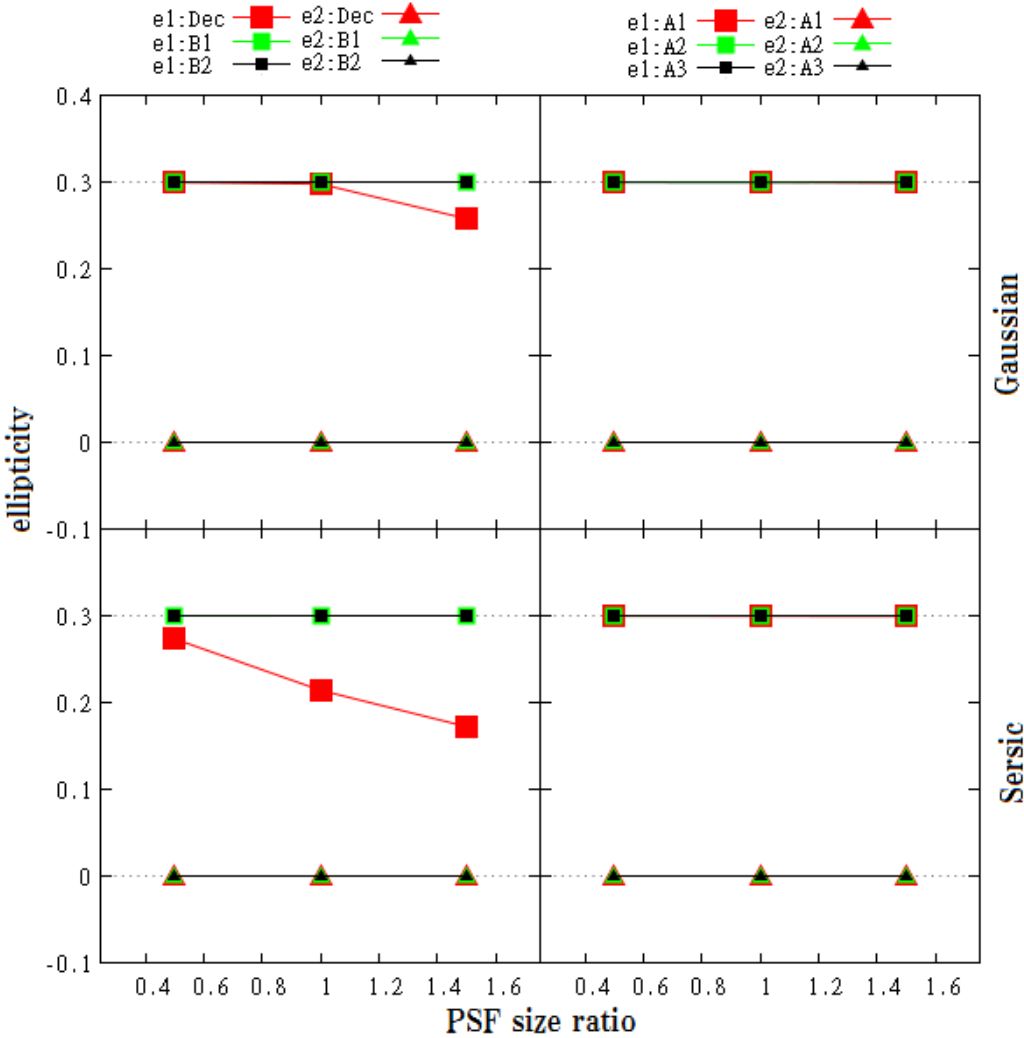}}
\caption{
\label{fig:sim_00}
The results of the PSF correction tests with circular PSF.
The differences between the four panels are the profile of the galaxy and the analysis method. Top panels are results with Gaussian galaxies and bottom panels are with S\'{e}rsic galaxies.  
Left panels are with methods: 
red symbols (Dec) indicate PSF correction by deconvolution,
green symbols (B1) indicate Method B1, and
black symbols (B2) indicate Method B2. Right panels are with methods: 
red symbols (A1) indicate PSF correction using Method A1,
green symbols (A2) indicate use of Method A2, and
black symbols (A3) indicate use of Method A3.
In each panel, the PSF size ratio (0.5, 1.0, 1.5)  is plotted on the horizontal axis and ellipticity with PSF correction is plotted on the vertical axis.
Squares indicate ellipticity 1 for which the true value is 0.3, and 
triangles indicate ellipticity 2 for which the true value is 0.0.
}
\end{figure*}

\begin{figure*}[htbp]
\centering
\resizebox{\hsize}{!}{\includegraphics{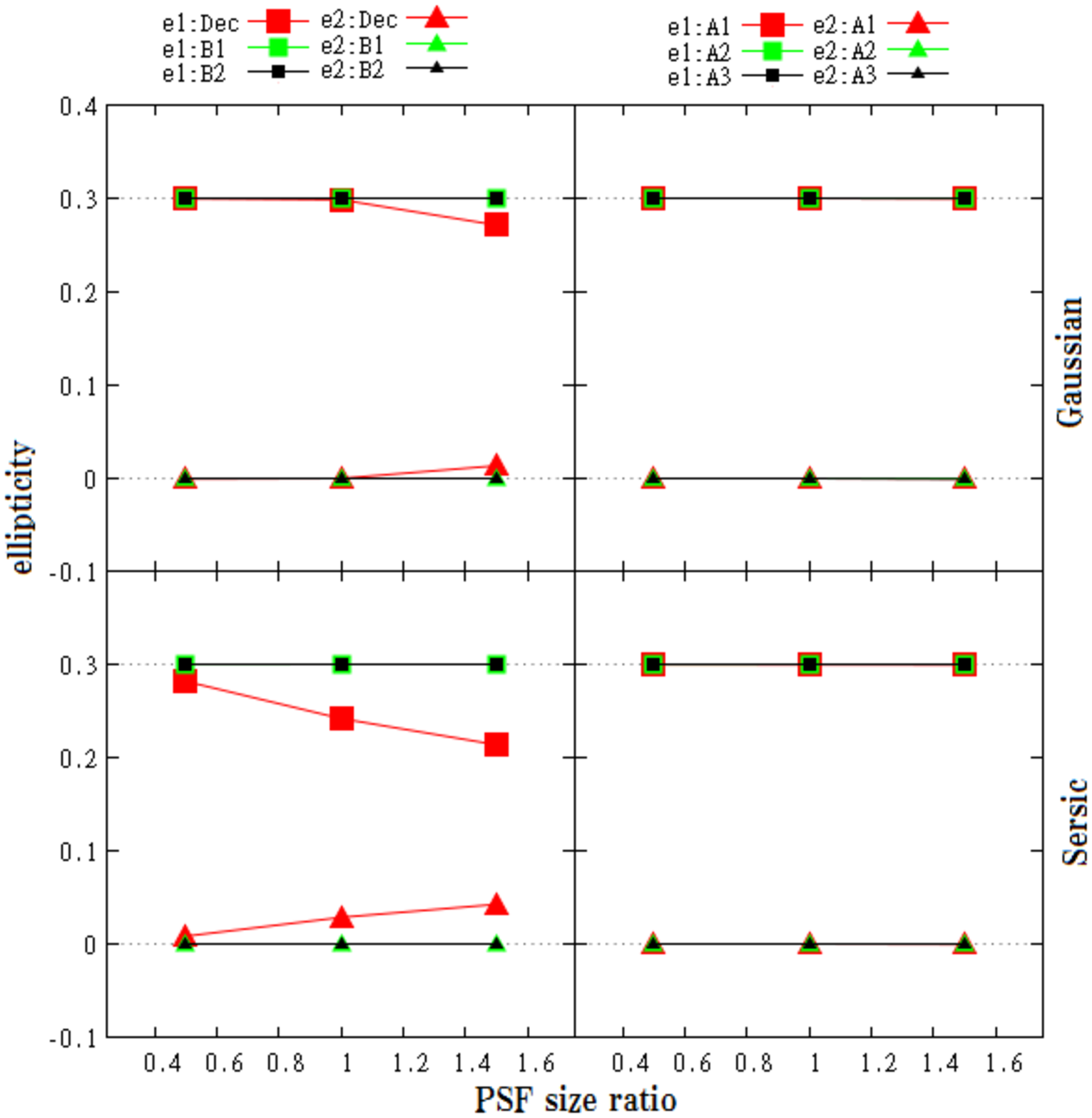}}
\caption{
\label{fig:sim_11}
This is the same as Figure \ref{fig:sim_00}, but PSF has ellipticity (0.1, 0.1).
}
\end{figure*}

\begin{figure*}[htbp]
\centering
\resizebox{\hsize}{!}{\includegraphics{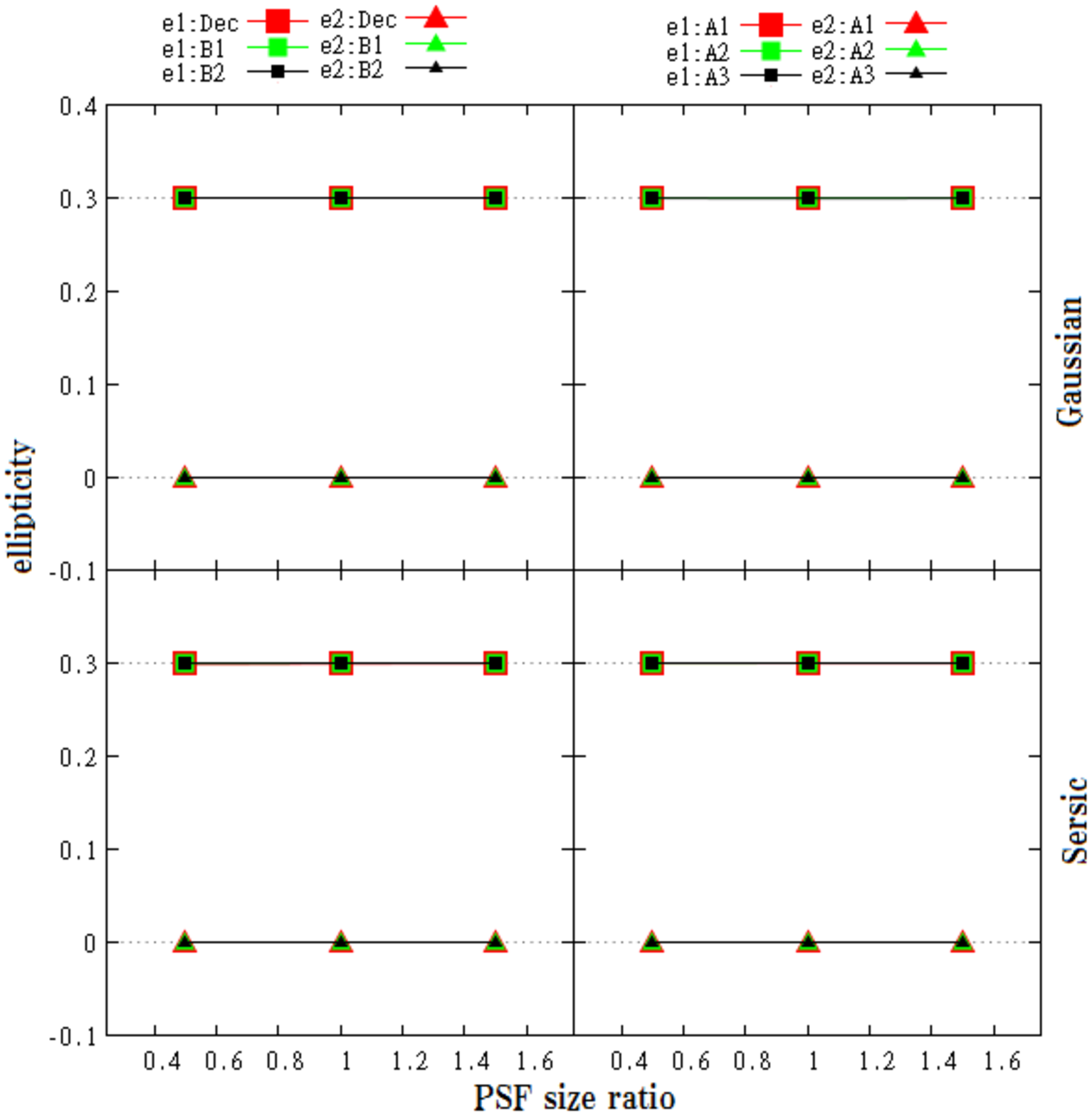}}
\caption{
\label{fig:sim_30}
This is the same as Figure \ref{fig:sim_00}, but PSF has ellipticity (0.3, 0.0).
}
\end{figure*}

\begin{figure*}[htbp]
\centering
\resizebox{\hsize}{!}{\includegraphics{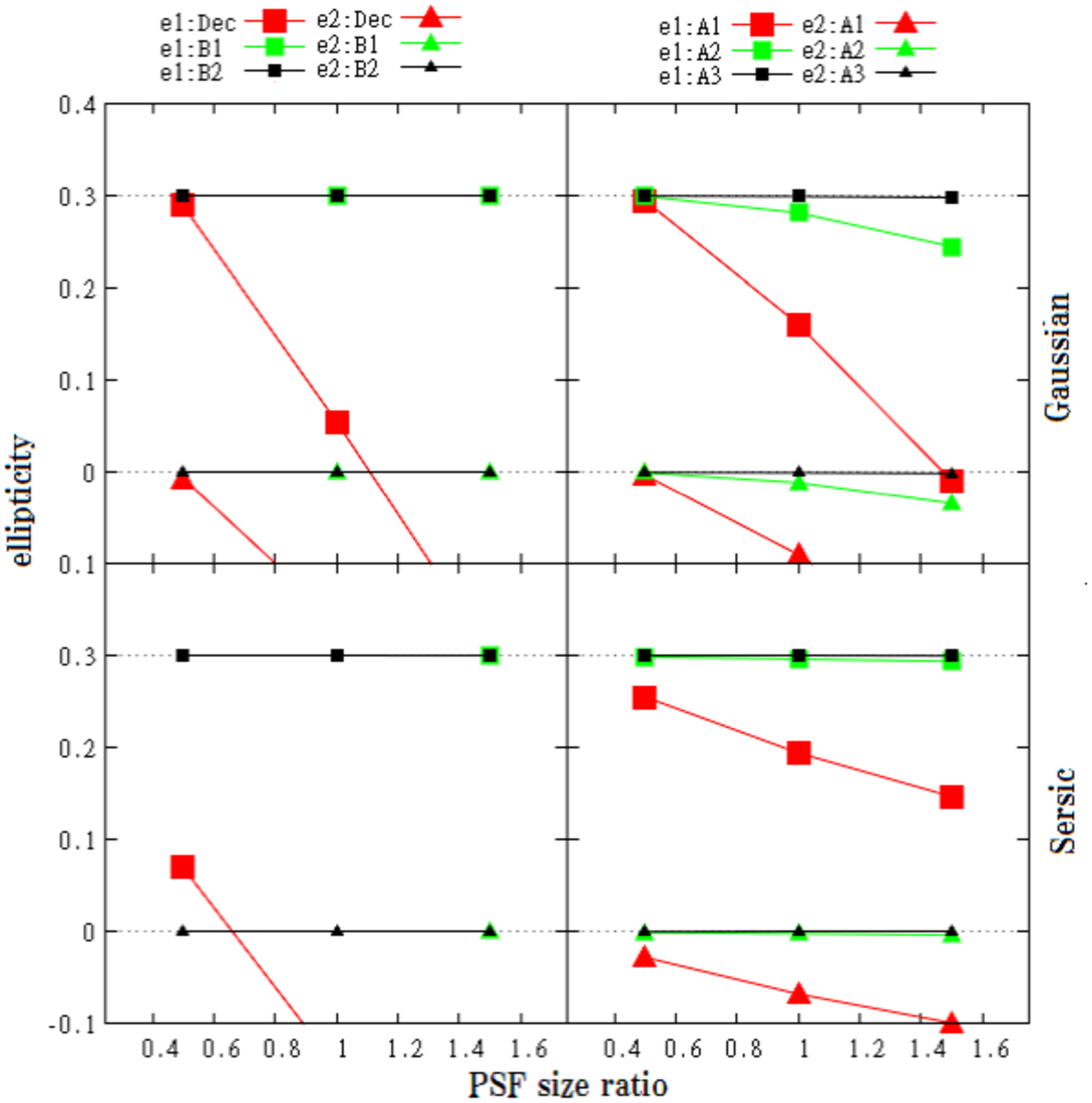}}
\caption{
\label{fig:sim_66}
This is the same as Figure \ref{fig:sim_00}, but PSF has ellipticity (-0.6, -0.6).
}
\end{figure*}

Figures \ref{fig:sim_00} to \ref{fig:sim_66} show the PSF corrected ellipticities.
The following list shows the simulation parameters for each figure set,
\\\hspace{-50pt}
\begin{tabular}{|c|c|c|c|c|c|c|c|c|}\hline
Figure number
 & \ref{fig:sim_00} & \ref{fig:sim_00}
 & \ref{fig:sim_11} & \ref{fig:sim_11}
 & \ref{fig:sim_30} & \ref{fig:sim_30}
 & \ref{fig:sim_66} & \ref{fig:sim_66}
\\\hline
Image Type 
 & G & G & G & G & G & G & G & G 
\\\hline
PSF Ellipticity 
 & 0 & 0 & 1 & 1 & 3 & 3 & 6 & 6
\\\hline
Method 
 & B & A & B & A & B & A & B & A
\\\hline\hline
Figure number
 & \ref{fig:sim_00} & \ref{fig:sim_00}
 & \ref{fig:sim_11} & \ref{fig:sim_11}
 & \ref{fig:sim_30} & \ref{fig:sim_30}
 & \ref{fig:sim_66} & \ref{fig:sim_66}
\\\hline
Image Type 
 & S & S & S & S & S & S & S & S
\\\hline
PSF Ellipticity 
 & 0 & 0 & 1 & 1 & 3 & 3 & 6 & 6
\\\hline
Method 
 & B & A & B & A & B & A & B & A
\\\hline
\end{tabular}\\

where Image Type [G, S] means using the [Gaussian, S\'{e}rsic] image;
PSF Ellipticity [0,1,3,6] means the ellipticity of the PSF is [(0.0, 0.0), (0.1, 0.1), (0.3, 0.0), (-0.6, -0.6)]; and 
Method [A, B] means PSF correction with Method [A, B, and simple deconvolution].
Figures \ref{fig:sim_00} and  \ref{fig:sim_11} show the results using a circular or small elliptical(0.1, 0.1) PSF.
These results show that the deconvolution method cannot correct the PSF, but the 
0-ERA method can correct a small elliptical PSF with no systematic error
(where no systematic error means that the systematic error is under 0.1\%).
Figure \ref{fig:sim_30} shows the cases with an intermediate elliptical PSF.
It can be seen that the 0-ERA method can correct the PSF, 
because the ERA method is formulated in such a way that the PSF has the same ellipticity with the objects.
Figure \ref{fig:sim_66} shows the results with a large elliptical PSF.
These results show that the 0-ERA method can correct the PSF with no systematic error
if appropriate parameters are chosen for the re-smearing function.
Similar results were obtained using the ERA method (2nd-ellipticity), which can be seen in the previous paper describing ERA.
Thus, the 0-ERA method can analyse weak lensing shear with an accuracy similar to the ERA method.

Table \ref{tab:result} is the PSF corrected ellipticities of the simulation with circular PSF.
\begin{table}[htbp]
\begin{tabular}{|c|c|c|c|}\hline
Method & Ratio=0.5 & Ratio=1.0 & Ratio =1.5
\\\hline
Deconvolution & 0.300, 0.000 & 0.298, 0.000 & 0.258, 0.000 
%\\\hline
%KSB &                  0.310, 0.000 & 0.308, 0.000 & 0.306, 0.000
\\\hline
ERA A1 &                0.300, 0.000 & 0.300, 0.000 & 0.299, 0.000
\\\hline
ERA A2 &                0.300, 0.000 & 0.300, 0.000 & 0.300, 0.000
\\\hline
ERA A3 &                0.300, 0.000 & 0.300, 0.000 & 0.300, 0.000
\\\hline
ERA B1 &                0.300, 0.000 & 0.300, 0.000 & 0.300, 0.000
\\\hline
ERA B2 &                0.300, 0.000 & 0.300, 0.000 & 0.300, 0.000
\\\hline
\end{tabular}
\caption{\label{tab:result}
The $\epsilon_{0th1}$ and $\epsilon_{0th2}$ of corrected PSF with  each PSF size ratio}
\end{table}

%----------------------------------------------------------------------------------------------------
%----------------------------------------------------------------------------------------------------
%----------------------------------------------------------------------------------------------------
%----------------------------------------------------------------------------------------------------
%----------------------------------------------------------------------------------------------------
%----------------------------------------------------------------------------------------------------
%----------------------------------------------------------------------------------------------------
%----------------------------------------------------------------------------------------------------
%----------------------------------------------------------------------------------------------------

%----------------------------------------------------------------------------------------------------
%----------------------------------------------------------------------------------------------------
%----------------------------------------------------------------------------------------------------
\section{Conclusions and comments}
We developed a new method of weak lensing analysis by defining a new ellipticity 
that is defined from 0th order moments.

This ellipticity is measured in the region near the centre, rather than the usual ellipticity which is defined 
from the  quadrupole moments.
Therefore, the new ellipticity has a higher S/N than that of quadrupole ellipticity,
and it is expected that this new method can measure weak gravitational shear with higher accuracy.
The gain in the S/N is about $\sqrt{2}\sim\sqrt{3}$ if the image is an elliptical Gaussian.
We measured the 0th-ellipticities and S/N using objects from Abell1689, and   
showed that the 0th- and 2nd-ellipticities have a high correlation.  
The gain in S/N using real data is almost the same as when using a simple elliptical Gaussian image 
and the gain in the number count using the 0th-ellipticity is about 1.5$\sim$2.
As a result, objects that are fainter  can be used in this new weak gravitational lensing 
shear analysis, 
and therefore more accurate shear analysis is expected. 
The mass reconstruction for galaxy clusters using this scheme will be
evaluated in  forthcoming papers.

We tested the PSF correction using this method in a simple simulation,
and the results of the simulation show that this new method is able to
correct  the PSF effect with enough accuracy by using an appropriate re-smearing function. 
In this simulation, we used a large image to neglect the noise
from pixelization, and therefore this test is not realistic. We  also did not 
consider pixel noise and the interpolation problem in PSF correction.
We succeeded in correcting PSF correction without any systematic error, 
but there are still many issues for precise weak lensing analysis in real situations, for examples in correcting complex PSF, PSF interpolation, stacking image, pixel noise, pixelization, and so on.  For example, we studied pixel noise effects on measuring ellipticity in Okura and Futamase 2013, but the pixel noise effect correction with PSF correction is very important for precise weak lensing analysis. All of these issues are very important and complicated problems, and thus need to be studied individually. 
The detailed evaluation of the systematic errors will be done in future studies.

%----------------------------------------------------------------------------------------------------
%----------------------------------------------------------------------------------------------------
%----------------------------------------------------------------------------------------------------
\begin{acknowledgements}
We would like to thank Alan T. Lefor for improving the English usage in this paper.
\end{acknowledgements}
%----------------------------------------------------------------------------------------------------
%----------------------------------------------------------------------------------------------------

%----------------------------------------------------------------------------------------------------
%----------------------------------------------------------------------------------------------------
%----------------------------------------------------------------------------------------------------
\begin{appendix}
\section{Ellipticity of the 0th order moment and the reduced shear}
\label{AP:0thE}
In this Appendix, we show detailed calculations for deriving the relation between the 0th-ellipticity 
and the reduced shear, and how to use the 0th-ellipticities $\mbeZ$ and $\bepZ$.

Since the zero image has circular image, the following equations are obtained,
\begin{eqnarray}
\label{eq:ZEROZERO1}
0&=&\int d^2\beta\frac{\bbe^2_2}{\bbe^2_0+\mbg^*\bbe^2_2}f(|\bbe|)
\nonumber\\&&=
\int d^2\beta\frac{\bbe^2_2}{\bbe^2_0+\mbg\bbe^{2*}_2}\IZt W^0(\bbe,0)\hspace{30pt}g<1\\
\label{eq:ZEROZERO2}
0&=&\int d^2\beta\frac{\bbe^2_2}{\bbe^2_0+\bbe^{2*}_2/\mbg^*}f(|\bbe|)
\nonumber\\&&=
\int d^2\beta\frac{\bbe^2_2}{\bbe^2_0+\bbe^{2*}_2/\mbg^*}\IZt W^0(\bbe,0)\hspace{20pt}g>1
,\end{eqnarray}
where $f$ is an arbitrary function, and $W^0$ is a Gaussian weight function, which is a function of 
the absolute value of $\bbe$ and the following approximation:
\begin{eqnarray}
\IZt&\approx&I^0(|\bbe|)
%\\
%W^0(\bbe,0)&=&W^0(|\bbe|,0).
.\end{eqnarray}

In the image plane, these functions are written as 
\begin{eqnarray}
\frac{\bbe^2_2}{\bbe^2_0+\mbg^*\bbe^2_2}&=&\frac{\bbe^1_1}{\bbe^{1*}_1+\mbg^*\bbe}\propto \frac{\bth^1_1-\mbg\bth^{1*}_1}{\bth^{1*}_1}=\frac{\bth^2_2}{\bth^2_0}-\mbg\\
\frac{\bbe^2_2}{\bbe^2_0+\bbe^2_2/\mbg}&=&\frac{\bbe^1_1}{\bbe^{1*}_1+\bbe^1_1/\mbg}\propto\frac{1}{\mbg} \frac{\bth^1_1-\bth^{1*}_1/\mbg}{\bth^1_1}=\mbg^2 \lr{\frac{1}{\mbg}-\frac{\bth^{2*}_2}{\bth^2_0}}\\
\IZt&=&\ILnsdt \\
W^0(\bbe,0)&=&exp\left({-\frac{\bbe^2_0}{\sigma_0^2}} \right)
 = exp \left({-\frac{\bth^2_0-\Real{\frac{2\mbg}{1+|\mbg|^2}\bth_2^{2*}}}{\sigma^2}} \right)
\nonumber\\
 &=& exp \left({-\frac{\bth^2_0-\Real{\bde^*\bth_2^2}}{\sigma^2}} \right) = W(\bth,\bde) = W(\bth,\bep_W),
\end{eqnarray}
where we temporarily set $\bep_W=\bde$.
Therefore, in the source plane eq. \ref{eq:ZEROZERO1} and eq. \ref{eq:ZEROZERO2} change to 
\begin{eqnarray}
0&=&\int d^2\theta\lr{\frac{\bth^2_2}{\bth^2_0}-\mbg}\ILnsdt W(\bth,\bde)=\lrs{\cZ^0_2-\mbg \cZ^0_0}_{(\ILnsd,\bde)}\nonumber\\&&\hspace{175pt}g<1\\
0&=&\int d^2\theta\lr{\frac{\bth^2_2}{\bth^2_0}-1/\mbg^*}\ILnsdt W(\bth,\bde)=\lrs{\cZ^0_2- \cZ^0_0/\mbg^*}_{(\ILnsd,\bde)}\nonumber\\&&\hspace{175pt}g>1
,\end{eqnarray}
and finally we obtain
\begin{eqnarray}
\label{eq:ZEROSOURCE1}
\lr{\frac{\cZ^0_2}{\cZ^0_0}}_{(\ILnsd,\bde)}&\equiv\mbe_{0th}&=\mbg\hspace{50pt}g<1\\
\label{eq:ZEROSOURCE2}
\lr{\frac{\cZ^0_2}{\cZ^0_0}}_{(\ILnsd,\bde)}&\equiv\mbe_{0th}&=\frac{1}{\mbg^*}\hspace{42pt}g>1.
\end{eqnarray}
Then $\bep_W$ is derived as 

\begin{eqnarray}
\bep_W=\bde=\frac{2\mbg}{1+|\mbg|^2}=\frac{2\mbe_{0th}}{1+|\mbe_{0th}|^2}\equiv\bep_{0th}.
\end{eqnarray}

\end{appendix}

\end{document}